\documentclass[preprint,12pt,showpacs]{revtex4}

\usepackage[brazil, english]{babel}
\usepackage[utf8]{inputenc}
\usepackage{graphicx}
\usepackage{epsfig}
\usepackage{amssymb}
\usepackage{amsmath}

\graphicspath{%
    {converted_graphics/}
    {/}
}
\begin{document}

\title{Deep Inelastic Scattering in the Exponentially Small Bjorken Parameter Regime from the Holographic Softwall Model}
\author{Eduardo Folco Capossoli$^{1,2,}$}
\email[Eletronic address: ]{educapossoli@if.ufrj.br}
\author{Henrique Boschi-Filho$^{1,}$}
\email[Eletronic address: ]{boschi@if.ufrj.br} 
\affiliation{ $^1$Instituto de F\'{\i}sica, Universidade Federal do Rio de Janeiro, 21.941-972 - Rio de Janeiro-RJ - Brazil \\
 $^2$Departamento de F\'{\i}sica, Col\'egio Pedro II, 20.921-903 - Rio de Janeiro-RJ - Brazil}

\begin{abstract}
We use the AdS/CFT correspondence and the holographic softwall model to investigate the Deep Inelastic Scattering (DIS) in the exponentially small $x$ (Bjorken Parameter) regime. We calculate the corresponding structure functions for scalar fields. Using these results we studied the problem of the saturation line in the strong interactions. Our results are consistent with those achieved using other models.
\end{abstract}


\maketitle


\section{Introduction}

At the end of the 1960s there were performed at the Stanford Linear Accelerator Center (SLAC) a series of experiments in which electrons with energy above 21 giga electron-volts (GeV) were directed towards `` proton-targets " and as a result of this scattering were produced a variety of hadronic final states. This type of scattering is known as deep inelastic scattering (DIS). This scattering has immense importance in high energy physics, because was through it that was revealed the internal structure of protons.

This scattering is called deep because it occurs in high energy  and it is inelastic because  the resulting particles are different from the initial state. The figure $1$ shows a pictorial representation of a deep inelastic scattering between a lepton $\ell$, exchanging a virtual photon of momentum $q$ and one target hadron with momentum $P$. $X$ represents the various other hadrons that can be produced from the collision.

\begin{figure}[!ht] \label{dis}
  \centering
  \includegraphics[scale = 0.40]{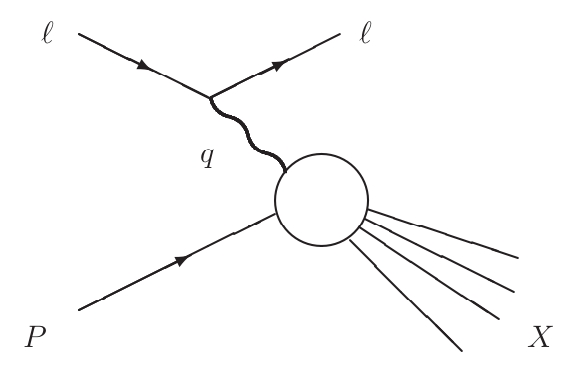} 
\caption{Deep inelastic scattering between a lepton and one target hadron.}
\end{figure}

The important parameters for the DIS are the virtuality of the photon, given by $q^2  = q_\mu q_\nu \eta^{\mu \nu} = -Q^2$, with metric signature $\eta^{\mu \nu} = (-, +, +, +)$, and the Bjorken variable $x$, defined by $x \equiv -q^2 /2.P.Q$

Some other parameters are also relevant, such as the mass $M$ of the initial hadron, such that $M^2 = - P^2$, and the squared center-of-mass energy, $s= -P^2_X = - (P + q)^2$. The condition $s \geq M^2$ defines the range of Bjorken variable $x$, such that $0\leq x \leq 1$. Strictly speaking, both for the DIS, and for other processes, the incoming electron corresponds to the $q^2 \rightarrow \infty$ limit with $x$ fixed. In the DIS case the Bjorken variable itself defines the inelasticity of scattering.

The differential cross section for DIS can be written as
\begin{equation}\label{f2}
d\sigma \propto \frac{\alpha^2}{q^4}L^{\mu \nu}_{electron} W_{\mu \nu}
\end{equation}

\noindent where $\alpha$ is the fine structure constant, $L^{\mu \nu}_{electron}$ is the leptonic tensor and $W_{\mu \nu}$ is known as the hadronic tensor and is the quantity of interest here. Assuming that the initial hadron state is not polarized, the hadronic tensor has the following form \cite{Manohar:1992tz}:
\begin{equation}\label{f3}
W^{\mu \nu} = i \int d^4 y~ e^{i q.y} \langle P, {\cal Q}|[J^{\mu}(y), J^{\nu}(0)]|P, {\cal Q} \rangle
\end{equation}

\noindent where $|P, ${\cal Q}$ \rangle$ represents a normalizable hadronic state with 4-momentum $P^{\mu}$  and electric charge ${\cal Q}$ of the initial hadron and $J^{\mu}$ is the electromagnetic hadronic current.

Using the gauge invariance, $q_{\mu}W^{\mu \nu} = 0$, and Lorentz covariance the tensor in (\ref{f3}) can be decomposed as follows:
\begin{equation}\label{f4}
W^{\mu \nu} = W_{1} \left( \eta^{\mu \nu} - \frac{q^{\mu} q^{\nu}}{q^2} \right) + \frac{2x}{q^2} W_2 \left( P^{\mu} + \frac{q^{\mu}}{2x} \right)\left( P^{\nu} + \frac{q^{\nu}}{2x} \right),
\end{equation}

\noindent where $W_1 (x, q^2)$ e  $W_2 (x, q^2)$ are called the structure functions. These functions describe the distribution of momenta of the quarks inside the protons.

For the particular regime where $x\sim 0.5$ and $q^2 \sim 1$ (GeV)$^2$ there is an approximate relation between these functions known as the Callan-Gross relation 
\begin{equation}\label{f5}
W_2 (x, q^2) \approx 2 x \;W_1 (x, q^2).
\end{equation}

Using the optical theorem one can relate the tensor $W^{\mu \nu}$ through the imaginary part of the forward Compton scattering, describe by the tensor:
\begin{equation}\label{f6}
T^{\mu \nu} = i \int d^4 y~ e^{i q.y} \langle P, {\cal Q}|T \left\{ J^{\mu}(y) J^{\nu}(0)\right\} |P, {\cal Q} \rangle, 
\end{equation}

\noindent where  $T\{{\cal O}_1 {\cal O}_2\}$ means that the operators product is temporally ordered.

As $T^{\mu \nu}$ has the same decomposition of $W^{\mu \nu}$, it also has the structure functions $\tilde{W}_1 (x, q^2)$ and  $\tilde{W}_2 (x, q^2)$. One can relate those structure functions, using the optical theorem, so:

\begin{equation}\label{f7}
W_s (x, q^2) \equiv 2 \pi ~\textrm{Im}~\tilde{W}_s (x, q^2); ~~~~~~s= 1,2. 
\end{equation}

The description of strong interactions via quantum chromodynamics (QCD) is consensus across the scientific community, even at low energy limit $ (g_ {YM}> 1) $ where the QCD can not be treated perturbatively.

An innovative proposal to overcome this difficulty is the anti-de Sitter correspondence/Conformal Field Theory (AdS/CFT) \cite{Maldacena:1997re, Gubser:1998bc, Witten:1998qj, Witten:1998zw, Aharony:1999ti}. This conjecture or duality relates a conformal ${\cal N} = 4$ super Yang-Mills theory (SYM) with symmetry group $SU(N)$, with $N \rightarrow \infty$, living in $3+1$ dimensional flat spacetime (Minkowski) with a superstring theory in a 10-dimensional curved spacetime known as anti de Sitter space which can be mathematically described as $AdS_{5}\times S^5$. From general arguments of the AdS/CFT conjecture one can see that there is an isomorphism between the Hilbert spaces of the bulk and boundary theories.

The SYM theory at the boundary  has a continuous spectrum of particles, which is a consequence of its conformal symmetry. Due to this symmetry, the AdS/CFT correspondence can not be used directly in the study of hadrons because in general one can not define an S matrix for this problem. Polchinski and Strassler \cite{Polchinski:2001tt} studied glueball scattering in four dimensions using the AdS/CFT correspondence and proposed that this scattering would correspond to a finite region of the $AdS_5$ space, thus the  conformal invariance should be broken. In the works \cite{BoschiFilho:2002ta, BoschiFilho:2002vd} a cut off at a certain value $z_{\rm max}$ of the radial $z$ coordinate of $AdS$ space was introduced and and it was considered a slice of AdS space in the region $0 \leq z \leq z_{\rm max}$ with a boundary condition at $z=z_{\rm max}$. Moreover one can associate the size of AdS slice with the energy scale of the QCD:
\begin{equation}\label{f10}
z_{max} = \frac{1}{\Lambda_{QCD}}.
\end{equation}

This model is known as the hardwall and although it has an abrupt infrared cut off it proved quite promising and some interesting results were found \cite{BoschiFilho:2005yh, deTeramond:2005su, Erlich:2005qh,Capossoli:2013kb}. For the DIS, within the hardwall context, some important results have been achieved \cite{Polchinski:2002jw,Brower:2006ea,Hatta:2007he, BallonBayona:2007rs, Gao:2009ze,Brower:2010wf}.

The softwall model (SW) arises with the aim to make the breaking of the conformal invariance with a softer infrared cut off. This model is quite interesting because it provides linear Regge trajectories for mesons and glueballs \cite{Karch:2006pv,Colangelo:2007pt}. These phenomenological models are known generically as AdS/QCD models. 

In this present work the structure functions will be calculated for the DIS using the formulation of AdS/QCD in the softwall model. Other studies for DIS have been discussed within the softwall model finding results consistent with the literature \cite{BallonBayona:2007qr, Braga:2011wa} and for other DIS studies using holographic models see for instance \cite{BallonBayona:2010ae, Cornalba:2008sp,Pire:2008zf,Albacete:2008ze, BallonBayona:2008zi, Yoshida:2009dw, Hatta:2009ra, Avsar:2009xf, Cornalba:2009ax, Bayona:2009qe, Cornalba:2010vk, Koile:2011aa, Koile:2013hba, Koile:2014vca, Koile:2015qsa}. 

In particular, in ref. \cite{BallonBayona:2007qr} the SW model was used to discuss the DIS problem in the large and small Bjorken parameters. Here, in this work, we extend this discussion to the exponentially small Bjorken parameter. This is important because the exponentially small $x$ regime is relevant to the discussion of saturation in QCD wich we will address here in the section V. This problem was discussed within the hardwall model in  \cite{Hatta:2007he}. Since the hardwall and softwall models have different spectra, in particular leading to different Regge trajectories, we want to study the saturation line in the softwall model to see if it leads to any different behavior.

This work is organized as follows: in section II, we review the main properties of the softwall model and its description of the DIS process. In section III, we discuss the exponentially small $x$ regime within the softwall model and calculate the corresponding structure functions. In the section IV, in order to emphasize our results found in section III, we compare then with the ones found in the literature. In section V, we use the results of the previous section to study the saturation line in strong interactions as implied by the softwall model. Finally, in section VI, we present our conclusions. 


\section{The softwall model and the DIS}


The holographic SW model was proposed originally for the study of vector mesons \cite{Karch:2006pv} and subsequently extended to glueballs \cite{Colangelo:2007pt}. The SW model introduces a decreasing exponential factor in the action for the fields by phenomenological motivation and to obtain a linear Regge trajectory. It is a natural extension of the hardwall model, that is, is an alternative way to break the conformal invariance in the boundary theory and, therefore, make it an approximate effective theory for QCD. In this model, the action for the fields found in the $AdS_5 \times S^5$ space will be \cite{BallonBayona:2007qr}:
\begin{equation}\label{acao_soft}
S = \int d^{10} x \sqrt{-g} \; e^{-\phi(z)} {\cal L}
\end{equation}

\noindent where ${\cal L}$ is the Lagrangean density, $\phi$ is a scalar dilaton field, $g$ is the determinant of the metric $g_{MN}$ of the $AdS_5 \times S^5$ space, given by:
\begin{equation}\label{gs}
ds^2 = g_{MN} dx^M dx^N= \frac{R^2}{z^2}(dz^2 + \eta_{\mu \nu}dy^\mu dy^\nu) + R^2 d\Omega^2_5, 
\end{equation}

\noindent $z$ is the holographic coordinate, $d\Omega^2_5$ is the angular measure on $S^5$, $R$ is the $AdS_5$ space radius given by $R = \alpha'^{1/2} \lambda^{1/4} = (4 \pi g_s N)^{1/4}\alpha'^{1/2}$, with $\alpha' \equiv l^2_s$ (Regge slope). In particular, in ref. \cite{Karch:2006pv} it was shown that if the dilaton is written as:
\begin{equation}\label{phi}
\phi(z) = k z^2
\end{equation}

\noindent then, one obtains linear Regge trajectories. 

This new approach does not change the geometry of $AdS_5$ space, that is, $0 \leq z \leq \infty $. However, it can be seen that the $e^{-k z^2}$ factor inserted in the action makes the integrand
becomes increasingly closer to zero as the value of $z$ increases, unlike the abrupt behavior of the hardwall model. The constant $k$ has dimension of mass  squared and is associated with the QCD scale.

In this work we use indices $M, N, \cdots$  to refer to the $10-$dimensional space, separating into $\mu, \nu, \cdots$ for the Minkowski spacetime at the boundary of $AdS_5$, $m, n, \cdots$ on $AdS_5$ and $a, b, \cdots$ on $S^5$. The quantities who live in $4-$dimensional spacetime will have their indices raised or lowered using the Minkowski metric $\eta^{\mu \nu} = (-1, +1, +1, +1)$. The indices $M, N, \cdots, m , n, \cdots, a, b, \cdots $ will be raised or lowered using the $10-$dimensional metric.

The ref. \cite{Polchinski:2002jw} showed, using the hardwall model, that the tensor given by  (\ref{f6}) can be decomposed as the sum intermediate states $X$ with mass $M_X$ produced in the collision, then:
\begin{eqnarray}\label{f11}
\textrm{Im}~T^{\mu \nu} & = & \pi \sum_{P_X, X} \langle P, {\cal Q}|J^{\nu}(0) |P_x, X \rangle \langle P_X, X | {\tilde J}^{\mu}(q)|P, {\cal Q} \rangle \nonumber \\ 
                        & = & 2 \pi^2 \sum_{X} \delta (M^2_X + [P + q]^2 ) \langle P, {\cal Q}|J^{\nu}(0) |P + q, X \rangle \langle P + q, X | J^{\mu}(0)|P, {\cal Q}\rangle
\end{eqnarray}

\noindent whose matrix element $ \langle P + q, X | J^{\mu}(0)|P, {\cal Q}\rangle$ represents the vertex of the interaction of the virtual photon of momentum $q$ with the initial hadron with momentum $P$ and the final hadron with momentum $P_X = P + q$. In the eq. (\ref{f11}) ${\tilde J}^{\mu}(q)$ represents the Fourier transform of $J^{\mu}(x)$. 

The vertices can be mapped to certain terms of interaction in supergravity. In the case of scalar hadrons the relevant interaction term to supergravity is associated with a Kaluza-Klein gauge field $A_m = (A_{\mu}, A_z )$, which does not depend on the coordinates of $S^5$ space, and two scalar fields $\Phi_i$ and $\Phi_X$ related respectively with the initial and final states of hadron. This relationship between the matrix element of the gauge theory and supergravity interaction term is presented in \cite{Polchinski:2002jw} , and is given by:
\begin{eqnarray}\label{f12}
\eta_{\mu}\langle P_X, X | {\tilde J}^{\mu}(q)|P, {\cal Q} \rangle & = & (2\pi)^4 \delta^4 (P_X - P - q) \eta_{\mu} \langle P + q, X | J^{\mu}(0)|P, {\cal Q}\rangle \nonumber \\
                                                                   & = & i {\cal Q} \int d^{10} x \sqrt{-g} A^m \left( \Phi_i \partial_m \Phi^{\ast}_X - \Phi^{\ast}_X \partial_m \Phi_i \right)
\end{eqnarray}

\noindent where $\eta_{\mu}$ is the photon polarization and the states $\Phi$ can be decomposed as
\begin{equation}\label{f13}
\Phi (y_{\mu},~ z,~ \Omega) = e^{iP.y}\psi(z, \Omega),
\end{equation}

\noindent a plane wave with $4-$momentun in the four physical dimensions times a function of the other six dimensions.

Here, in the SW case, there is a modification in (\ref{f12}) due the dilaton background field. Then (\ref{f12}) can be written as \cite{BallonBayona:2007qr}:
\begin{eqnarray}\label{f14}
\eta_{\mu}\langle P_X, X | {\tilde J}^{\mu}(q)|P, {\cal Q} \rangle & = & (2\pi)^4 \delta^4 (P_X - P - q) \eta_{\mu} \langle P + q, X | J^{\mu}(0)|P, {\cal Q}\rangle \nonumber \\
                                                                   & = & i {\cal Q} \int d^{10} x \sqrt{-g} \;e^{-k z^2} A^m \left( \Phi_i \partial_m \Phi^{\ast}_X - \Phi^{\ast}_X \partial_m \Phi_i \right)
\end{eqnarray}

An important observation is that the fields $A_m$ and $\Phi$ in (\ref{f14}) are not the same as the ones found in (\ref{f12}) since they have been modified by the introduction of dilatonic field. As a consequence, the solutions presented in (\ref{f14}) will involve the confluent hypergeometric functions instead of Bessel functions, as in (\ref{f12}).

The action for a massless gauge field in the SW model is given by:
\begin{equation}\label{f15}
S = -  \int d^{10} x \sqrt{-g} \; e^{-\phi(z)} \frac{1}{4} F^{m n} F_{m n}
\end{equation}

\noindent which leads to the following equation of motion
\begin{equation}\label{f16}
\partial_m [e^{-kz^2} \sqrt{-g}\; F^{m n}] = 0
\end{equation}

\noindent or explicitly:
\begin{eqnarray}\label{f17}
\Box A_{z}   & - &  \partial_z \left[\partial_{\mu} A^{\mu} \right]  = 0\nonumber \\ 
\Box A_{\mu} & + &  ze^{kz^2} \partial_z \left[ e^{-kz^2} \frac{1}{z} \partial_z A^{\mu} \right] - \eta^{\mu \beta} \partial_ {\beta}\left[z e^{kz^2} \partial_z \left[e^{-kz^2} \frac{1}{z} A_z\right] + \partial_{\nu}\left(\eta^{\nu \alpha} A_{\alpha} \right) \right] = 0,
\end{eqnarray}

\noindent with boudary condition
\begin{equation}\label{f18}
A_{\mu}(y_{\mu}, z \to 0) = \eta_{\mu} e^{iq.y}
\end{equation}

\noindent that represents the insertion of an operator $\eta_{\mu} {\tilde J}^{\mu}(q)$. The gauge field satisfies Maxwell's equation $D^{m} F_{m n} = 0$ in the $AdS_5$ space, where $D^{m}$ is the covariant derivative and $F_{m n} = \partial_m A_n - \partial_n A_m$. Choosing a suitable Lorentz-like gauge:
\begin{equation}\label{f19}
z\; e^{kz^2} \partial_z \left[e^{-kz^2} \frac{1}{z} A_z\right] + \partial_{\nu}\left(\eta^{\nu \alpha} A_{\alpha} \right)  = 0,
\end{equation}

\noindent the field equations are:
\begin{eqnarray}\label{f20}
A_z (y^{\mu},z)      & = &  {\mathbb A} \; e^{iq.y} \; z \; {\cal U}(1 + \frac{q^2}{4k}; 1; kz^2) \nonumber \\ 
A_{\mu}(y^{\mu},z)   & = &  {\mathbb B}_{\mu} \;e^{iq.y} \; z^2 \; {\cal U}(1 + \frac{q^2}{4k}; 2; kz^2)
\end{eqnarray}

\noindent where ${\cal U}(1 + \frac{q^2}{4k}; 1; kz^2)$ and ${\cal U}(1 + \frac{q^2}{4k}; 2; kz^2)$ are confluent hypergeometric functions of the second kind and ${\mathbb A}$ and ${\mathbb B}_{\mu}$ are parameters to be calculated in the following.

To find ${\mathbb B}_{\mu}$ we used the boundary condition (\ref {f18}) and the following relationship, with $z\rightarrow 0$, so that
\begin{equation}\label{f21}
{\cal U}(1 + \frac{q^2}{4k}; 2; kz^2) \to \frac{\Gamma(1)}{\Gamma(1 + \frac{q^2}{4k})} (kz^2)^{-1}.
\end{equation}

Then one gets ${\mathbb B}_{\mu} = \eta_{\mu}k \Gamma(1 + \frac{q^2}{4k})$. So $A^{\mu}$ is given by:
\begin{equation}\label{f22}
A_{\mu}(y^{\mu},z) =  \eta_{\mu}\; k \; \Gamma(1 + \frac{q^2}{4k}) \;e^{iq.y} \; z^2 \; {\cal U}(1 + \frac{q^2}{4k}; 2; kz^2).
\end{equation}

Using this results in (\ref{f17}), one gets ${\mathbb A} = \frac{i q. \eta}{2} \; \Gamma(1 + \frac{q^2}{4k})$ so that, $A_{z}$ is given by:
\begin{equation}\label{f23}
A_z (y^{\mu},z) =  \frac{i q. \eta}{2} \; \Gamma(1 + \frac{q^2}{4k}) \;e^{iq.y} \; z \; {\cal U}(1 + \frac{q^2}{4k}; 1; kz^2).
\end{equation}

Looking at the expressions (\ref{f22}) and  (\ref{f23}) one can realize that both potential decreases rapidly for the following situation: 
\begin{equation}\label{f24}
\left(1 + \frac{q^2}{4k}\right) kz^2 > 1.
\end{equation}

\noindent Therefore, one can define an effective maximum for the radial coordinate $z$, which is independent of cut off scale $ k $, then:
\begin{equation}\label{f25}
z_{\rm int} \approx \frac{1}{\sqrt{k \left(1 + \frac{q^2}{4k}\right)}} ~ \sim ~\frac{1}{q}.
\end{equation}

For the scalars fields from (\ref{acao_soft}), with ${\cal L} = \partial_m \Phi \partial^m \Phi + m^2_5 \Phi^2$, one has the action 
\begin{equation}\label{f26}
S = \int d^{10} x \sqrt{-g} \; e^{-\phi(z)} ( \partial_m \Phi \partial^m \Phi + m^2_5 \Phi^2)
\end{equation}

\noindent which leads to the equations of motion, up to a multiplicative constant factor
\begin{equation}\label{f27}
\partial_m [e^{-kz^2} \sqrt{-g}\; \partial^m \Phi]  - \sqrt{-g}e^{-kz^2}m^2_5 \Phi = 0.
\end{equation}

The introduction of mass $m^2_5$ is made to adjust the conformal dimension $\Delta$ of the field associated with the hadron. According to the holographic dictionary, the conformal dimension of on operator is related with a $p-$form by the following relationship: $R^2m^2_5  = (\Delta - p)(\Delta + p -4)$ \cite{Csaki:1998qr}. For a $0-$form, $R^2m^2_5  = \Delta(\Delta-4)$.

Making explicit the dependence on $z$ coordinate, one can write:
\begin{equation}\label{f28}
z^3 e^{-kz^2} \partial_z [e^{-kz^2} \frac{1}{z^3}\; \partial^z \Phi]  + \Box \Phi -  \frac{R^2}{z^2}m^2_5 \Phi = 0.
\end{equation}

Using the ansatz for the scalar field (\ref{f13}) and the following normalization conditions
\begin{equation}\label{f29}
\int dz\; d^5 \Omega \;\frac{R^8}{z^3} \sqrt{g_{\Omega}}\;e^{-kz^2} |\Phi|^2 = \int d^5 \Omega \;\sqrt{g_{\Omega}}\; |\psi|^2 = 1
\end{equation}

\noindent one can get normalizable solutions, given by:
\begin{equation}\label{f30}
\Phi (y_{\mu},~ z,~ \Omega) = \left[\frac{2k^{\Delta - 1}\Gamma (n+1)}{\Gamma(n + \Delta - 1)}\right]^{1/2} \frac{1}{R^4}\; e^{ip.y}\; z^{\Delta} \;L^{\Delta - 2}_{n} (kz^2)\; \psi(\Omega)
\end{equation}

\noindent where $L^m_n$ are the associated Laguerre polynomials with order $n$, with $n \geq 0 $.

As was shown in \cite{BallonBayona:2007qr} one can obtain the following relationship:
\begin{equation}\label{f32}
-n = \frac{p^2}{4k} + \frac{\Delta}{2}; \;\;\;\;(n= 0, 1,2 \cdots)
\end{equation}

\noindent where $p^2 = m^2_n$ are the masses of the hadronic states. Then, assuming that the initial hadron with momentum $p = P$ has the lowest mass in the  spectrum $(n=0)$ equals to $\sqrt{2k\Delta}$, it follows that the initial state of the hadron is given by:

\begin{equation}\label{f33}
\Phi_i (y_{\mu},~ z,~ \Omega) = \left[\frac{2k^{\Delta - 1}}{\Gamma( \Delta - 1)}\right]^{1/2} \frac{1}{R^4}\; e^{iP.y}\; z^{\Delta} \; \psi(\Omega)
\end{equation}

In the case of final state with momentum $p = P_X$, one can establish a similar relationship to (\ref {f32}) given by:
\begin{equation}\label{f34}
-n_X = \frac{P^2_X}{4k} + \frac{\Delta}{2} = -\frac{s}{4k} + \frac{\Delta}{2};\;\;\;\;(n_X= 0, 1,2 \cdots)
\end{equation}

\noindent where $P_X = P + q = -s$.

Thus, the final state of hadron can be written as follows:
\begin{equation}\label{f35}
\Phi_X (y_{\mu},~ z,~ \Omega) = \left[\frac{2k^{\Delta - 1} \Gamma(\frac{s}{4k} - \frac{\Delta}{2} + 1)}{\Gamma( \frac{s}{4k} + \frac{\Delta}{2} - 1)}\right]^{1/2} \frac{1}{R^4}\; e^{iP_X.y}\;z^{\Delta} \; L^{\Delta - 2}_{n_X}(kz^2)\;\psi(\Omega)
\end{equation}

As the last step in this section it is necessary to establish now how to represent the exponentially small $x$ regime that will be studied in this work. The energy scale for the DIS is given by, $s = - (P +q)^2 \approx q^2 \left( \frac{1}{x} - 1\right)$. The holographic relationship between the momenta in string theory and momenta in gauge theory is given by ${\tilde p_{\mu}} = p_{\mu}\;z/R$, where ${\tilde p}_{\mu}$ is the momentum as seen by an inertial observer placed in $10-$dimensions. The quantities for $4-$ dimensional gauge theory are written without tilde and quantities with tilde denote $ 10-$ dimensional objects. From this, one can write the following relationship:
\begin{equation}\label{f36}
\alpha ' {\tilde s} = \alpha s \frac{z^2}{R^2} + \nabla^2_z + \nabla_{\alpha} \nabla^{\alpha}  \stackrel{<}{\sim} \frac{q^2}{x} \frac{z^2}{(\alpha ' 4 \pi g_s N)^{1/2}}.
\end{equation} 

Assuming that the interactions occur effectively in the region $0 < z < z_{\rm{int}}$, with $z_{\rm{int}}$ given by (\ref{f25}), the above inequality implies that:
\begin{equation}\label{f37}
\alpha ' {\tilde s}  \stackrel{<}{\sim} \frac{(\alpha ' 4 \pi g_s N)^{-1/2}}{x} .
\end{equation} 

This relationship tells us that the DIS energy scale in the framework of string theory ${\tilde s}$ has a maximum determined by the Bjorken variable $x$. The regime to be studied in this work is defined as $x \sim \exp (- \sqrt{g_s N}) $.


\section{The exponentially small $x$ regime in the  softwall model}


The DIS in the exponentially small $x$ regime is characterized by multiple pomeron exchange represented by gravitons in de AdS/CFT correspondence \cite{Polchinski:2002jw}.

The dominant contribution at high energies to the strings scattering amplitude in the $10-$dimensional SW model is
\begin{equation}\label{f38}
S_{{\rm string}} = \int d^{10}x \sqrt{-g} \; e^{-\phi(z)} {\cal L}_{{\rm eff,string}}
\end{equation}

\noindent which is identified with the amplitude of forward Compton scattering in four dimensions with the following form:
\begin{equation}\label{f39}
\eta_{\mu} \eta_{\nu} T^{\mu \nu} (2 \pi)^4 \delta^4 (q - q') = S_{{\rm string}}.
\end{equation}

In (\ref{f38}) one can identify ${\cal L}_{{\rm eff,string}} = ({\cal K}G)|_{t=0}$, where ${\cal K}$ is a kinematic factor and $G$ is the Virasoro-Shapiro factor for closed strings in a $10-$dimensional Minkowski space, given by:

\begin{equation}\label{f40}
G= \frac{\alpha'^3{\tilde s}^2}{64} \displaystyle\prod_{{ \tilde \xi} = \tilde s, \tilde t,  \tilde u}  \frac{\Gamma(-\alpha' \tilde \xi / 4)}{\Gamma(1 + \alpha' \tilde \xi / 4)}.
\end{equation}

Then, one can write (\ref{f38}) as
\begin{eqnarray}\label{f41}
S_{{\rm string}} & = & \frac{1}{8} \int d^{10}x \sqrt{-g} \; e^{-\phi(z)} \Bigl\{ 4 v^a v_a \partial_m \Phi F^{mn} F_{pn} \partial^p \Phi \nonumber \\
                 & - & \left( \partial^M \phi \partial_M \Phi v^a v_a + 2 v^a \partial_a \Phi v^b \partial_b \Phi \right) F^{mn} F_{mn} \Bigl\} G|_{t=0}
\end{eqnarray}

\noindent where $v^a$ are the Killing vectors of the compact space $S^5$ (or generically $W$), $F^{mn}$ is associated with an incoming photon with $4-$momentum $q_{\mu}$ and outgoing with $4-$momentum $q'_{\mu}$ and $\Phi$ represents the incoming and outgoing scalars state with $4-$momentum $P^{\mu}$ and $P^{\mu}_X$, respectively . 

The condition $x < < 1$ implies $P \cdot q >> q^2 >> P^2$, then the dominant contribution comes from the first term in (\ref{f41}) corresponding to $(P \cdot q)^2$ when $m = \mu$ and $p = \nu$. Then:
\begin{equation}\label{f46}
S_{{\rm string}}  =  \frac{1}{2} \int d^{10}x \sqrt{-g} \; e^{-\phi(z)} \; v^a v_a \;\partial_{\mu} \Phi(-P) \; \partial^{\nu} \Phi(P) \; F^{\mu n}(-q') \; F_{\nu n}(q)  ~ G|_{t=0}
\end{equation}

Using the Stirling approximation for the Virasoro-Shapiro factor and performing an expansion around $\tilde t = 0$ the expression (\ref {f40}) can be written as:
\begin{equation}\label{f42}
G= - \left[ \frac{1}{\tilde t} + \pi \; {\rm cot} \left( \frac{\pi \alpha' {\tilde s}}{4} \right) \right] \left[ \frac{\alpha' {\tilde s}}{4e} \right]^{\alpha' {\tilde t}^2 / 2} [1 + {\cal O} (\alpha' {\tilde t}^2)]
\end{equation}

As ${\cal K}$ is real, the imaginary part of $S_{{\rm string}}$ is determined by the imaginary part of $ G|_{\tilde t \to 0}$. Then the imaginary part from excited strings is
\begin{equation}\label{f43}
{\rm Im}G|_{\tilde t \to 0} =   \frac{\pi \alpha'}{4} \displaystyle\sum_{m=1}^{\infty} \delta (m - \alpha' {\tilde s}/4) (m)^{\alpha' {\tilde t}/2}.
\end{equation}

It is important to note that in (\ref{f43}) the last factor, i.e., the small-angle Regge behavior of the string amplitude, becomes important for ultra small $x$ since the variable $ \tilde t $ in ten dimensions is not exactly zero even if $ t = 0$. This occurs because $\tilde{t}$ includes in addition derivatives in the transverse direction just like $s$ as can be seen below in Eq. (\ref{f44}). Although $\tilde{t}$ is small on the string scale, of order $(g_s N)^{-1/2}$, its effects becomes large for exponentially large  $s$ or equivalently to  exponentially small $x$.

Following the prescription given in \cite{Polchinski:2002jw} by inserting ${\tilde s}^{\alpha' {\tilde t}/2} \sim x^{-\alpha' {\tilde t}/2}$, averaging over the delta functions in (\ref{f43}), since $\tilde s \propto 1/x$ as can be seen in (\ref{f37}), and replacing $\tilde t $ as a differential operator acting on the $t$ channel, then:
\begin{equation}\label{f71}
{\rm Im}G|_{\tilde t \to 0} =   \frac{\pi \alpha'}{4} \displaystyle\sum_{m=1}^{\infty} \delta (m - \alpha' {\tilde s}/4) (x)^{{-\alpha' \nabla^2}/2}.
\end{equation}

The Mandelstam variables in the $10-$dimensional space, $\tilde t$ e $\tilde s$, are related to the variables $ t $ and $ s $ of $4-$dimensional space by
\begin{eqnarray}\label{f44}
\alpha' {\tilde s} & = & \alpha' s \frac{z^2}{R^2} + \frac{\alpha'}{R^2} \left( -3 z \partial_z + z^2 \partial^2_z + \nabla^2_W \right) \nonumber \\
\alpha' {\tilde t} & = & \alpha' t \frac{z^2}{R^2} + \frac{\alpha'}{R^2} \left( -3 z \partial_z + z^2 \partial^2_z + \nabla^2_W \right) 
\end{eqnarray}

Assuming that everything is smooth in the radial direction, therefore, $z \partial_z = {\cal O}(1)$  and the dimensionless Laplacian on $W$  is assumed to be similar to ${\cal O}(1)$, then, using (\ref{f44}) one gets $\alpha' {\tilde s}  =  \alpha' s ({z^2}/{R^2})$ plus corrections of order ${\alpha'}/{R^2} \sim (g_s N)^{-1/2}$ which for excited strings states can be neglected compared to the integer $m$ in the delta function. Following this reasoning one can write (\ref {f43}) as follows:
\begin{equation}\label{f45}
{\rm Im}G|_{\tilde t \to 0} \approx   \frac{\pi \alpha'}{4} \displaystyle\sum_{m=1}^{\infty} \delta (m - \frac{ \alpha' s~ z^2}{4 R^2}) (x)^{{-\alpha' \nabla^2}/2}.
\end{equation}

Finally, one can write (\ref{f46}) as
\begin{eqnarray}\label{f47}
{\rm Im} \;S_{{\rm string}} &  = & \frac{\pi \alpha' }{8} \displaystyle\sum_{m=1}^{\infty} \int d^{10}x \sqrt{-g} \; e^{-\phi(z)} \; v^a v_a \nonumber \\ 
                 &    & \times \;\partial_{\mu} \Phi(-P) \; \partial^{\nu} \Phi(P) \; F^{\mu n}(-q') \; F_{\nu n}(q)  ~ \delta (m - \frac{ \alpha' s~ z^2}{4 R^2}) (x)^{{-\alpha' \nabla^2}/2}.
\end{eqnarray}

The fields strengths will be calculated as solutions of the gauge fields given in (\ref{f22}) and (\ref{f23}), so that
\begin{eqnarray}\label{f48}
F_{0 \mu} & = & \partial_0 A_{\mu} - \partial_{\mu} A_0 = \partial_u A_{\mu} - \partial_{\mu} A_u  \nonumber \\ 
          & = & \eta_{\mu} e^{iqy} \;\Gamma \left(1 + \frac{q^2}{4k}\right) \partial_z \left[k z^2 \;{\cal U} \left(1 + \frac{q^2}{4k}; 2; kz^2 \right) \right]  \nonumber \\ 
          &   & - (i q_{\mu}) \; i \frac{q . \eta}{2} \;\Gamma \left(1 + \frac{q^2}{4k}\right) e^{iqy}\; k \; {\cal U} \left(1 + \frac{q^2}{4k}; 1; kz^2 \right).
\end{eqnarray}

Making the following substitutions: $X = k z^2$, $z = \sqrt{X/k}$ e $\partial_z = 2 \;k^{1/2} \;X^{1/2} \partial_X$, then:
\begin{eqnarray}\label{f51}
 F_{0 \mu} & = & \eta_{\mu} e^{iqy} \; \Gamma \left(1 + \frac{q^2}{4k}\right) [2 k z]\; \partial_X \left [ X \; {\cal U} \left(1 + \frac{q^2}{4k}; 2; X \right) \right] \nonumber \\ 
          &   & +  q_{\mu} \; \frac{q . \eta}{2} \;\Gamma \left(1 + \frac{q^2}{4k}\right) e^{iqy}\; k \; {\cal U} \left(1 + \frac{q^2}{4k}; 1; kz^2 \right)
\end{eqnarray}

\noindent Using the following property of confluent hypergeometric functions: 
\begin{equation}
x \;\partial_x \;{\cal U}(a,b,x) = (1-b)\; {\cal U}(a,b,x) - (1+a-b)\;{\cal U}(a,b-1,x) \nonumber
\end{equation}

\noindent one can write (\ref{f51})
\begin{equation}\label{f52}
F_{0 \mu}  =  e^{iqy} \; {\cal U} \left(1 + \frac{q^2}{4k}; 1; kz^2 \right) \;\Gamma \left(1 + \frac{q^2}{4k}\right) \frac{z}{2} \; \left[-q^2 \eta_{\mu} + q_{\mu} (q . \eta) \right]\,.
\end{equation}

For the second field strength:
\begin{eqnarray}\label{f53}
F_{\mu \nu} & = & \partial_{\mu} A_{\nu} - \partial_{\nu} A_{\mu} = i q_{\mu} A_{\nu} - i q_{\nu} A_{\mu}  \nonumber \\ 
          & = & i\; e^{iqy} \;{\cal U} \left(1 + \frac{q^2}{4k}; 2; kz^2 \right) \;\Gamma \left(1 + \frac{q^2}{4k}\right) k z^2 \left[q_{\mu} \eta_{\nu} - q_{\nu} \eta_{\mu} \right]   
\end{eqnarray}

Using (\ref{f52}) and (\ref{f53}) one can compute:
\begin{eqnarray}\label{f54}
F^{0 \mu}(-q') & = & g^{00} g^{\mu \nu} F_{0 \nu} (-q')   \nonumber \\ 
               & = & \frac{z^2}{R^2} \frac{z^2}{R^2} \eta^{\mu \nu} F_{0 \nu} (-q')  .
\end{eqnarray}

\noindent regrouping some terms one finds
\begin{equation}\label{f60}
F^{\mu n} (-q') F_{\nu n}(q)  = e^{i(q - q')} \mathbb{G}^{\mu}_{\nu}(q, q')
\end{equation}

\noindent by defining
\begin{eqnarray}\label{f59}
\mathbb{G}^{\mu}_{\nu}(q, q') & \equiv &  \frac{z^4}{R^4}\; \Gamma \left(1 + \frac{q'^2}{4k}\right) \Gamma \left(1 + \frac{q^2}{4k}\right) \Biggl\{ \frac{z^2}{4}{\cal U} \left(1 + \frac{q'^2}{4k}; 1; kz^2 \right) {\cal U} \left(1 + \frac{q^2}{4k}; 1; kz^2 \right) \nonumber \\
           &        & \times \left[-q'^2 \eta_{\mu} + q^{\nu} (q . \eta) \right] \left[-q^2 \eta_{\nu} + q_{\nu} (q . \eta) \right] + \;k^2 z^4 \nonumber \\
           &        & \times \; {\cal U} \left(1 + \frac{q'^2}{4k}; 2; kz^2 \right){\cal U} \left(1 + \frac{q^2}{4k}; 2; kz^2 \right) \left[q'^{\mu} \eta^{\gamma} - q'^{\gamma} \eta^{\mu} \right] \left[q_{\nu} \eta_{\gamma} - q_{\gamma} \eta_{\mu} \right] \Biggl\} 
\end{eqnarray}

For the scalars states one can use the solution to initial states showed in (\ref{f33}) and is given by:
\begin{equation}\label{f61}
\Phi_i (P) = a\; e^{iP.y}\; z^{\Delta} \; \psi(\Omega)
\end{equation}

\noindent where $a = \left[\frac{2k^{\Delta - 1}}{\Gamma( \Delta - 1)}\right]^{1/2} (1/R^4) $, so that
\begin{eqnarray}\label{f62}
\partial_{\mu} \Phi(-P) \partial^{\nu} \Phi(P) & = & -i P_{\mu} \;a \;e^{iPy} \;z^{\Delta} \;\psi^{\ast}(\Omega) \; \frac{z^2}{R^2}\; i P^{\nu}\; a\; e^{iPy} \;z^{\Delta} \;\psi(\Omega) \nonumber \\
                                               & = & a^2 \; \frac{z^{2 \Delta + 2}}{R^2} \;|\psi(\Omega)|^2 \; P_{\mu} P^{\nu} \;\Phi(-P) \Phi(P) \nonumber \\
                                               & = & a^2 \; \frac{z^{2 \Delta + 2}}{R^2} \;|\psi(\Omega)|^2 \; P_{\mu} P^{\nu} \;\Phi^{\ast} \Phi
\end{eqnarray}

By replacing (\ref{f60}) and (\ref{f62}) in (\ref{f47}), one gets:
\begin{eqnarray}\label{f63}
{\rm Im } \;S_{{\rm string}} &  = & \frac{\pi \alpha' }{8} \displaystyle\sum_{m=1}^{\infty} \int d^{4}y \; dz \; d^5 \Omega \; \frac{R^{10}}{z^5}\; \sqrt{-g_W} \; e^{-kz^2} \; v^a v_a \;a^2 \; \frac{z^{2 \Delta + 2}}{R^2} \;|\psi(\Omega)| \; P_{\mu} P^{\nu} \nonumber \\ 
                 &    & \times \; e^{i(q - q')y } \; \mathbb{G}^{\mu}_{\nu}(q, q') ~ \delta (m - \frac{ \alpha' s~ z^2}{4 R^2}) (x)^{{-\alpha' \nabla^2}/2} \;\Phi^{\ast} \Phi .
\end{eqnarray}

\noindent The nabla operator in the exponent is indeed the Laplacian which describes diffusion into the fifth dimension for the spin 2 exchanged graviton. This operator appears due to the fact that there is a finite amount of momentum transferred into the $z$ direction, although the momentum transferred in four physical dimensions is strictly zero \cite{Polchinski:2002jw, Brower:2006ea}.

According to \cite{Hatta:2007he}, one can write:
\begin{equation}\label{f1000}
\nabla^2 \equiv \left(\frac{R}{z}\right)^2 \nabla_0^2 \left(\frac{R}{z}\right)^{-2},
\end{equation}

\noindent with $\nabla_0^2 = (1 /\sqrt{|g|}) \; \partial_z [\sqrt{|g|} \; g^{zz} \;\partial_z]$.

For the following redefinition $\beta \equiv \ln (z^2_0/z^2)$ with $z = QR^2$ and $  z_0 = \Lambda R^2$, one gets:
\begin{equation}\label{f1001}
\nabla^2  = \frac{4}{R^2}(\partial_{\beta}^2 - 1),
\end{equation}

\noindent then
\begin{equation}\label{f1002}
(x)^{{-\alpha' \nabla^2}/2} \;(\Phi_i^{\ast} \Phi_i) = (x)^{{\alpha' \xi}/2} \;\Phi_i^{\ast} \Phi_i, ~~~~\xi = \frac{4}{R^2}(1 - \Delta^2)
\end{equation}

\noindent where we used (\ref{f1001}) and the fact that $\Phi_i \sim z^{\Delta} \sim e^{- \Delta\;\beta }$, as was showed in (\ref{f61}). 

We will use the following normalization,
\begin{equation}\label{f64}
\int d^5 \Omega \sqrt{-g_W} \; v^a v_a |\psi(\Omega)|^2 = \rho R^2
\end{equation}

\noindent where $\rho$ is a dimensionless quantity. 

Integrating (\ref{f63}) over $d^4y$ and using (\ref{f1002}), one can write
\begin{eqnarray}\label{f66}
{\rm Im } \;S_{{\rm string}} &  = & \frac{\pi \alpha' \rho R^{10}(x)^{{\alpha' |\xi|}/2}}{8} \; a^2 \; P_{\mu} P^{\nu}  (2 \pi)^4 \delta (q - q')  \nonumber \\
                             &    & \times \displaystyle\sum_{m=1}^{\infty} \int  \; dz\; e^{-kz^2}\; z^{2 \Delta - 3}\; \mathbb{G}^{\mu}_{\nu}(q, q') ~ \delta (m - \frac{ \alpha' s~ z^2}{4 R^2}).
\end{eqnarray}

Expanding the delta function in (\ref {f66}) as follows:
\begin{equation}\label{f68}
\delta (m - \frac{ \alpha' s~ z^2}{4 R^2}) =  \delta (f(z)) = \sum_{a_n}^{\infty} \frac{1}{|f'(a_n)|} \delta (z - a_n)  = \frac{1}{|f'(z_m)|} \delta (z - z_m)
\end{equation}

\noindent then
\begin{equation}\label{f70}
\delta (m - \frac{ \alpha' s~ z^2}{4 R^2}) =  \frac{2 R^2}{\alpha' s z_m} \delta (z - z_m).
\end{equation}

One can note that using (\ref{f36}), $z_m$, given by  $z_m = 2R \sqrt{m/\alpha's}$ can be approximated as:
\begin{equation}\label{f72}
z_m \approx \frac{2}{q}\; (4 \pi g_s N)^{1/4} \;(mx)^{1/2}
\end{equation} 

Using (\ref{f59}) and (\ref{f70}) in (\ref{f66}), one gets:
\begin{eqnarray}\label{f76}
{\rm Im } \;S_{{\rm string}} &  = & (2 \pi)^4 \delta (q - q') \;\eta_{\mu} \eta{\nu} \;\frac{\pi \rho R^{8}(x)^{{\alpha' |\xi|}/2}}{4s} \; a^2 \;  \Gamma^2 \left(1 + \frac{q^2}{4k}\right) \nonumber \\
                             &    & \times \;\Biggl\{ \displaystyle\sum_{m=1}^{\infty} e^{-kz^2_m}\;  \frac{z^{2 \Delta + 2}_m}{4} \; {\cal U}^2 \left(1 + \frac{q^2}{4k}; 1; kz^2_m \right) (-q^2)  \left[p^{\mu} - \frac{p.q}{q^2} q^{\mu} \right] \left[p^{\nu} - \frac{p.q}{q^2} q^{\nu} \right]  \nonumber \\
                             &    & + \; k^2 \; \displaystyle\sum_{m=1}^{\infty}\;  e^{-kz^2_m}\; z^{2 \Delta + 4}_m\; {\cal U}^2 \left(1 + \frac{q^2}{4k}; 2; kz^2_m \right) (p.q)^2 \nonumber \\
                             &    & \times \;\left[\eta^{\mu \nu} - \frac{(p^{\nu}q^{\nu} +  p^{\nu} q^{\mu})}{p.q} + \frac{q^2}{(p.q)^2} p^{\mu} p^{\nu} \right] \Biggl\}.
\end{eqnarray}

With (\ref{f39}) and (\ref{f72}), one obtains:
\begin{eqnarray}\label{f81}
{\rm Im } \;T^{\mu \nu} &  = & \frac{\pi \rho R^{8}(x)^{{\alpha' |\xi|}/2}}{4s} \; a^2 \; \frac{q^4}{4 x^2} \Biggl\{ \left[\eta^{\mu \nu} - \frac{q^{\mu}q^{\nu}}{q^2}  \right] \;{\cal I}_2 \nonumber \\
                        &    & + \; \left[p^{\mu} + \frac{q^{\mu}}{2x} \right] \left[p^{\nu} + \frac{q^{\nu}}{2x} \right] 4x^2  \left( {\cal I}_1 + \frac{{\cal I}_2}{q^2} \right) \Biggl\} 
\end{eqnarray}

\noindent where
\begin{eqnarray}\label{f78}
{\cal I}_1 &  \equiv & \frac{1}{4} \; \Gamma^2 (j) \;\displaystyle\sum_{m=1}^{\infty} e^{-kz^2_m}\;  z^{2 \Delta + 2}_m \; {\cal U}^2 \left(j ; 1; kz^2_m \right) \\
{\cal I}_2 &  \equiv & k^2 \; \Gamma^2 (j) \; \displaystyle\sum_{m=1}^{\infty}\;  e^{-kz^2_m}\; z^{2 \Delta + 4}_m\; {\cal U}^2 \left(j ; 2; kz^2_m \right) \label{f666}
\end{eqnarray}

\noindent  by defining  $j \equiv 1 + \frac{q^2}{4k} $.

Using  (\ref{f81}) with $a = \left[\frac{2k^{\Delta - 1}}{\Gamma( \Delta - 1)}\right]^{1/2} (1/R^4)$, one can write the structure functions (\ref{f7}) as:
\begin{eqnarray}\label{f82}
W_1 &  = & \frac{\pi^2 \rho k^{\Delta - 1}}{4 \Gamma (\Delta - 1)} \; \frac{q^4 (x)^{{\alpha' |\xi|}/2}}{s x^2} \;{\cal I}_2 \nonumber \\
W_2 &  = & \frac{\pi^2 \rho k^{\Delta - 1}}{4 \Gamma (\Delta - 1)} \; \frac{q^4 (x)^{{\alpha' |\xi|}/2}}{s x^2}  \;(2 x q^2) \left( {\cal I}_1 + \frac{{\cal I}_2}{q^2} \right)
\end{eqnarray}

\noindent 
These are the structure functions for DIS within the softwall model in the exponentially small $x$ regime. This is the main result of this section. 

In the next section we are going to compare this result with others known from the literature. In section V, we are going to use this result to discuss the saturation line in large $N_c$ QCD. 


\section{Comparison of the results}


In this section we will to compare ours structure functions, $W_1$ and $W_2$, represented by Eq. (\ref{f82}) found in section III, using the holographic softwall model with the ones obtained by Polchinski and Strassler in ref. \cite{Polchinski:2002jw} within the holographic hardwall model.

In order to do this comparison, we will rewrite the expressions of ${\cal I}_1$ and ${\cal I}_2$, given by Eqs. (\ref{f78}) and (\ref{f666}) since the structure functions, $W_1$ and $W_2$, are dependent of these expressions.

So, let us start using the following property for the confluent hypergeometric functions of the second kind:
\begin{equation}\label{f85}
\displaystyle\lim_{\stackrel{j \to +\infty}{(q^2 \to +\infty)}} {\cal U} \left( j; b; \frac{\zeta_m}{j - 1}\right)  = \displaystyle\lim_{j \to +\infty} \frac{2}{\Gamma (1 + j - b)} \zeta_m^{(1- b)/2} K_{b- 1} (2 \sqrt{\zeta_m})
\end{equation}

\noindent with $b = 1, 2 $,  $\zeta_m \equiv (4 \pi g_s N)^{1/2} m x$ and $K_{\ell-1}(\Gamma)$ are the modified Bessel functions. Then $\zeta_m = (j - 1) k z^2_m = \frac{q^2 z_m^2}{4} < q^2 z^2_{\rm int}$.

For ${\cal I}_1$, given by (\ref{f78}), one has
\begin{equation}\label{f89}
\displaystyle\lim_{\stackrel{j \to +\infty}{(q^2 \to +\infty)}} {\cal I}_1  =  \left[ \frac{q^2}{4}\right]^{- \Delta - 1}  \; \frac{\Gamma^2 (j)}{\Gamma^2 (j)} \;\displaystyle\sum_{m=1}^{\infty} \zeta_m^{\Delta + 1}\;  K^2_0 (2 \sqrt{\zeta_m})
\end{equation}

\noindent and then
\begin{equation}\label{f90}
 {\cal I}_1  \approx  \left[ \frac{q^2}{4}\right]^{- \Delta - 1}  \; \displaystyle\sum_{m=1}^{\infty} \zeta_m^{\Delta + 1}\;  K^2_0 (2 \sqrt{\zeta_m}).
\end{equation}

In a similar way for ${\cal I}_2$, using $\Gamma (j) = (j - 1) \Gamma (j-1)(x)^{{\alpha' |\xi|}/2}
$, one gets
\begin{equation}\label{f96}
 {\cal I}_2  \approx  4 \left[\frac{q^2}{4}\right]^{-\Delta} \;\displaystyle\sum_{m=1}^{\infty} \zeta^{\Delta + 1}\; K^2_1 (2 \sqrt{\zeta_m}).
\end{equation}

One can further approximate the series in (\ref{f90}) and (\ref{f96}) by integrals with the following substitutions, as showed below.

For ${\cal I}_1$, for instance, one has: 
\begin{equation}\label{f100}
{\cal I}_1  \approx  \left[ \frac{q^2}{4}\right]^{- \Delta - 1}  \; \displaystyle\sum_{m=1}^{\infty} \zeta_m^{\Delta + 1}\;  K^2_0 (2 \sqrt{\zeta_m}) \approx  \left[ \frac{q^2}{4}\right]^{- \Delta - 1} \int_1^{\infty} dm \left( \frac{w^2_m}{4} \right)^{\Delta + 1} K^2_0(w_m)
\end{equation}

\noindent or
\begin{equation}\label{f101}
  {\cal I}_1 \approx \left[ \frac{q^2}{4}\right]^{- \Delta - 1} \frac{4^{-\Delta - 1}}{(4 \pi g_s N)^{1/2} \;2 x}\int_{w_1}^{\infty} dw_m \;w^{2\Delta + 3}_m \; K^2_0(w_m)
\end{equation}

\noindent where $w_m \equiv 2 \sqrt{\zeta_m}$  and  $w_1 = 2 (4 \pi g_s N)^{1/4} \sqrt{x}$.

Now defining:
\begin{equation}\label{f102}
 {\tilde {\cal I}}_{0,\;2\Delta + 3} \equiv \int_{w_1}^{\infty} dw_m \;w^{2\Delta + 3}_m \; K^2_0(w_m)
\end{equation}

\noindent then one can write
\begin{eqnarray}\label{f103a}
 {\tilde {\cal I}}_{0,\;2\Delta + 3} &  =    & \int_{0}^{\infty} dw_m \;w^{2\Delta + 3}_m \; K^2_0(w_m) - \int_{0}^{w_1} dw_m \;w^{2\Delta + 3}_m \; K^2_0(w_m)  \\
                                     &  =    & {\cal I}_{0,\;2\Delta + 3} - [2 (4 \pi g_s N)^{1/4} \sqrt{x}]^{2 \Delta + 4}\; \frac{\ln^2 [2 (4 \pi g_s N)^{1/4} \sqrt{x}\;]}{2 \Delta + 4}.\label{f103}
\end{eqnarray}

Now one can get a suitable expression for ${\cal I}_1$, given by:
\begin{equation}\label{f1011}
  {\cal I}_1 \approx  \frac{q^{-2\Delta - 2}}{(4 \pi g_s N)^{1/2} \;2 x}\left[  {\cal I}_{0,\;2\Delta + 3} - [2 (4 \pi g_s N)^{1/4} \sqrt{x}]^{2 \Delta + 4}\; \frac{\ln^2 [2 (4 \pi g_s N)^{1/4} \sqrt{x}\;]}{2 \Delta + 4} \right]     
\end{equation}

This equation represents the difference between the structure functions at exponentially small $x$ calculated from the softwall model to the hardwall model. In fact, ${\tilde {\cal I}}_{0,\;2\Delta + 3}$, 
Eq. (\ref{f103a}), gives the contribution from the softwall model, while ${\cal I}_{0,\;2\Delta + 3}$ represents the contribution from the hardwall model. So, the difference on the structure functions resides on the second term in the r.h.s. of Eq. (\ref{f103}). 
Since in this work the regime of interest is the exponentially small $x$, it is clear that the second term in (\ref{f103}) becomes negligible, then it follows that:
\begin{equation}\label{ap}
{\tilde{{\cal I}}}_{0, \;2\Delta +3} \cong {\cal I}_{0, \;2\Delta +3}
\end{equation}

\noindent and the structure functions of the two models are approximately equal at very small $x$. Then Eq. (\ref{f90}) can be written as
\begin{equation}\label{f104}
 {\cal I}_1  \approx   \frac{q^{- 2\Delta - 2}}{(4 \pi g_s N)^{1/2} \;2 x}  \; {\cal I}_{0,\;2\Delta + 3}
\end{equation}

\noindent which is essentially the hardwall result \cite{Polchinski:2002jw}. 

Analogously for ${\cal I}_2$ one can approximate:
\begin{equation}\label{f105}
{\cal I}_2  \approx  4 \left[\frac{q^2}{4}\right]^{-\Delta} \displaystyle\sum_{m=1}^{\infty} \zeta_m^{\Delta + 1}\;  K^2_1 (2 \sqrt{\zeta_m}) \approx \frac{q^{- 2\Delta}}{(4 \pi g_s N)^{1/2} \;2 x}\int_{w_1}^{\infty} dw_m \;w^{2\Delta + 3}_m \; K^2_1(w_m)
\end{equation}

\noindent by defining
\begin{equation}\label{f106}
 {\tilde {\cal I}}_{1,\;2\Delta + 3} \equiv \int_{w_1}^{\infty} dw_m \;w^{2\Delta + 3}_m \; K^2_1(w_m)
\end{equation}

\noindent then
\begin{eqnarray}\label{f107}
 {\tilde {\cal I}}_{1,\;2\Delta + 3} &  =    & \int_{0}^{\infty} dw_m \;w^{2\Delta + 3}_m \; K^2_1(w_m) - \int_{0}^{w_1} dw_m \;w^{2\Delta + 3}_m \; K^2_1(w_m) \nonumber \\
                                     &  =    & {\cal I}_{1,\;2\Delta + 3} - \frac{[2 (4 \pi g_s N)^{1/4} \sqrt{x}\;]^{2 \Delta + 2}}{2 \Delta + 2} \label{diff2}\\
 {\tilde {\cal I}}_{1,\;2\Delta + 3} & \cong & {\cal I}_{1,\;2\Delta + 3}                         
\end{eqnarray}

\noindent In the above approximation we used the same reasoning as that for obtaining Eq. (\ref{ap}). 
The equation (\ref{diff2}) represents the difference between the soft and hardwall models for the structure function $W_2$. The second term in Eq. (\ref{diff2}) is explicitly this difference and is very small for very small $x$. 
Then, Eq. (\ref{f96}) can be written as
\begin{equation}\label{f108}
 {\cal I}_2  \approx   \frac{q^{- 2\Delta }}{(4 \pi g_s N)^{1/2} \;2 x}  \; {\cal I}_{1,\;2\Delta + 3}
\end{equation}

\noindent where
\begin{equation}\label{f109}
{\cal I}_{r,s} \equiv \int_{0}^{\infty} dw \;w^s \; K^2_r(w) = 2^{(s - 2)} \; \frac{\Gamma(\frac{s+1}{2} + r)\; \Gamma(\frac{s+1}{2} - r) \; \Gamma^2(\frac{s+1}{2})}{\Gamma (s+1)}
\end{equation}

With these approximations one can rewrite the softwall structure functions at exponentially small $x$ regime given by (\ref{f82}), as
\begin{eqnarray}\label{f112}
W_1 (x, q^2)  & \approx & \frac{\pi^2 \rho \; (x)^{{-2 + \alpha' |\xi|}/2}}{8 \;(4 \pi g_s N)^{1/2}\; \Gamma (\Delta - 1)}  \left(\frac{k}{q^2}\right)^{\Delta - 1} \; {\cal I}_{1,\;2\Delta + 3} 
\\
W_2 (x, q^2) & \approx & \frac{\pi^2 \rho \; (x)^{{-1 + \alpha' |\xi|}/2}}{4 \;(4 \pi g_s N)^{1/2}\; \Gamma (\Delta - 1)} \left(\frac{k}{q^2}\right)^{\Delta - 1} \;\left({\cal I}_{0,\;2\Delta + 3} + {\cal I}_{1,\;2\Delta + 3}  \right).\label{f115}
\end{eqnarray}

Computing the ratio of these structure functions one finds
\begin{equation}\label{f116}
\frac{ W_2 (x, q^2)}{W_1 (x, q^2)}   \approx  2x \; \frac{{\cal I}_{0,\;2\Delta + 3} + {\cal I}_{1,\;2\Delta + 3} }{{\cal I}_{1,\;2\Delta + 3}}.
\end{equation}

Using ${\cal I}_{0,n} = \frac{n-1}{n+1} \, {\cal I}_{1, n}$
\begin{equation}\label{f117}
{\cal I}_{0,2 \Delta + 3} = \frac{2 \Delta + 2}{2 \Delta + 4 } {\cal I}_{1, 2 \Delta +3}
\end{equation}

\noindent so that, one has:
\begin{equation}\label{f118}
\frac{ W_2 (x, q^2)}{W_1 (x, q^2)}   \approx  2x \; \left( \frac{2 \Delta +3 }{\Delta + 2} \right) 
\end{equation}

One can note that this ratio  is the same found in \cite{Polchinski:2002jw} using the hardwall model. There is a small difference between the structure functions in the soft and hardwall models. This difference can be seen for instance in Eq. (\ref{f103}). The first term in the r.h.s. of this equation gives exactly the contribution that is equal to the one from the hardwall model. Then, the second term in Eq. (\ref{f103}) gives the difference between the two models. But this term is proportional to $x^{2\Delta +4}$ and since $x$ is exponentially small, this difference is also exponentially small. 

\section{The saturation line in the softwall model}

Another question that can be analyzed from the DIS in the exponentially small $x$ regime within the softwall model, concerns the saturation line in QCD. This study was done previously in \cite{Hatta:2007he} within the hardwall model. 

The saturation is an interesting issue in QCD, because in a QCD phase diagram, $\tau - \beta$, where $\tau = \ln 1/x$ and $\beta = \ln Q^2/\Lambda^2 $, the saturation line represents a transition between weak and strong scattering regions. The saturation line can also be understood, for a fixed virtuality $Q$,  as a transition between linear and non-linear regimes.

The saturation line is related to a saturation scale, $Q_s(x)$, where $Q_s(x)$ is the saturation momentum, which can be roughly express as $Q_s^2(x) \sim x^{- \lambda}$, where $\lambda$ is the saturation exponent. 
The Figure $2$ summarizes some important informations about the saturation in QCD.

\begin{figure}[!ht] \label{mapa}
  \centering
  \includegraphics[scale = 0.40]{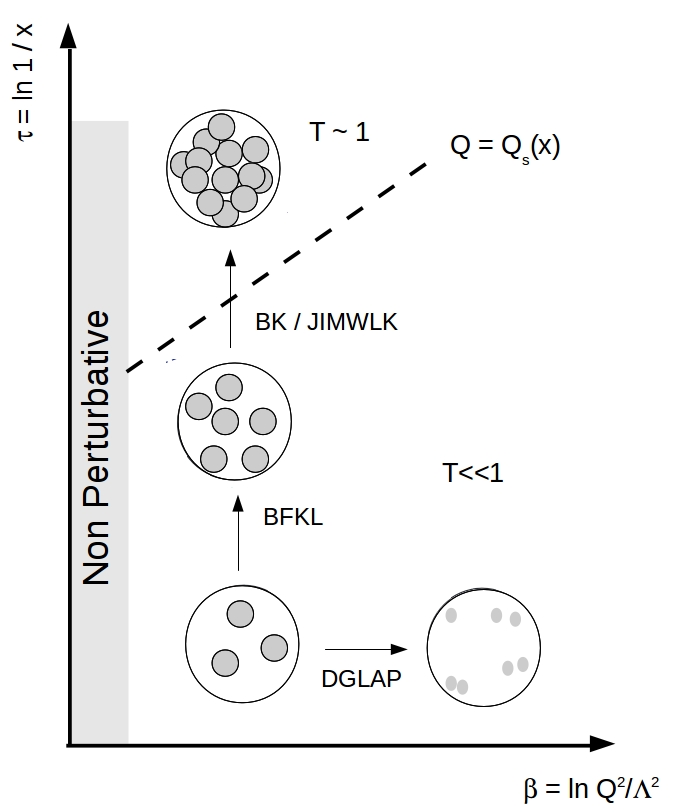} 
\caption{Map of high energy QCD.}
\end{figure}

The dashed line represents the saturtation line and the region bellow to this line represents the region where the scattering amplitude $(T)$ is weak, i.e,  $T<< 1$. The above region represents the strong scattering amplitude region, i.e, $T \simeq 1$. The scattering amplitude $T$ will be defined in Eq. (\ref{tb}). 

Looking at the Figure $2$, one can see in a schematic form the typical configuration of the protons (great circles) in different positions of the phase diagram. Considering initially for a low energy and low virtuality, the proton can be seen as formed by three partons (small circles).

For a fixed energy and increasing the virtuality or resolution $Q^2$, the number of partons increases while the size of gluons decreases, in fact, the number of partons increases logarithmically while the size of gluons decreases as $\sim 1/Q^2$, so that the proton becomes more and more diluted. This evolution is described in QCD by Dokshitzer-Gribov-Lipatov-Altarelli-Parisi or DGLAP equations \cite{Gribov:1972ri,DGLAP,DGLAP2}.

On other hand, for a fixed $Q^2$ and increasing the energy one can note a quickly growing in the nunber of partons, but keeping the same size, resulting in an increase in density inside the proton. This evolution is described in QCD by a linear equation, known as the Balitisky-Kuraev-Fadin-Lipatov or BFKL equation \cite{Lipatov:1976zz,BFKL,BFKL2}.

If the energy is still increasing (fixed $Q^2$) one has to take into account non-linear effects which are described by the Balistisky-Kovchegov or BK equation  \cite{Kovchegov:1999yj,BK}, and the Jalilian-Marian-Iancu-McLerran-Weigert-Leonidov-Kovner  or JIMWLK equation\cite{JalilianMarian:1997jx,JIKLMW2,JIKLMW3,JIKLMW4,JIKLMW5,JIKLMW6,JIKLMW7,JIKLMW8}. 

After this quick revision, let us start our analysis performing a change of variables in the metric given by (\ref{gs}), such that:
\begin{equation}
u = \frac{R^ 2}{z}; \; \; \;  \;0 \leq u \leq \infty
\end{equation}

\noindent then
\begin{equation}\label{us}
ds^2 = g_{MN} = \frac{R^2}{u^2} du^2 + \frac{u^ 2}{R^ 2} (\eta_{\mu \nu}dy^\mu dy^\nu) + R^2 d\Omega^2_5.
\end{equation}

Moreover, from the expression (\ref{f63}) it is clear that the structure function is proportional to the term shown below:
\begin{equation}\label{dif}
W_2(x, q^2) \propto (x)^{\alpha' \nabla^ 2/2}\Phi^{\ast} \Phi.
\end{equation}

\noindent  
Such term, will be in fact the relevant one for the discussion of the saturation line.

Using the following substitution, $\tau \equiv  \ln{(1/x)}$, where the rapidity $\tau$ can be interpreted as the evolution time, one can rewrite (\ref{dif}), with the following spectral decomposition:
\begin{equation}
\left( \frac{1}{x} \right)^{1 + \frac{\alpha' \nabla^ 2}{2}} \Phi^{\ast}(u) \Phi(u) \equiv \frac{1}{x} \int du' \langle u|e^ {\frac{\alpha' \tau \nabla^ 2}{2}}|u\rangle |\Phi(u')|^2
\end{equation}

Using (\ref{f1001}), with $\beta \equiv \ln (u^2/u^2_0)$, where $u = QR^2$ and $  u_0 = \Lambda R^2$, then, one gets:
\begin{equation}
e^{\tau + \frac{\alpha' \tau}{2} \left[ \frac{4}{R^2} (\partial^2_{\beta}-1)\right]} \Phi^{\ast}(u) \Phi(u) \equiv \frac{1}{x} \int du' \langle u|e^ {\frac{\alpha' \tau \nabla^ 2}{2}}|u\rangle |\Phi(u')|^2.
\end{equation}

Then using the integral representation of the heat kernel we have:
\begin{equation}\label{difu}
e^{\tau + \frac{2 \alpha' \tau}{R^2} (\partial^2_{\beta}-1) }\Phi^{\ast}(\beta) \Phi(\beta) = \int d\beta' \int \frac{d \nu}{2 \pi} e^{ i \nu (\beta - \beta' )} e^{\tau - \frac{2 \tau}{\sqrt{\lambda}}(\nu^2 + 1)} \;|\Phi(\beta')|^2
\end{equation}

\noindent where $\lambda \equiv R^2/\alpha' $ .

Solving the Gaussian integral in (\ref{difu}), one gets: 
\begin{equation}\label{difu_1}
e^{\tau + \frac{2 \alpha' \tau}{R^2} (\partial^2_{\beta}-1) }\Phi^{\ast}(\beta) \Phi(\beta) = \frac{e^{\omega_0 \tau}}{2 \sqrt{\pi D \tau}} \int_{0}^{\infty} d\beta'   e^{- \frac{(\beta - \beta' )^2}{4 D \tau}} e^{- \Delta \beta'}
\end{equation}

\noindent where we defined $\omega_0 = 1 - 2/\sqrt{\lambda}$, $D= 2/\sqrt{\lambda}$, and  we  also used the fact that $|\Phi(\beta')|^2 \sim u^{-2 \Delta} \sim e^{-\Delta \beta'}$. 
The equation (\ref{difu_1}) shows a Pomeron-like particle with an intercept $1 + \omega_0 = 2 - 2/\sqrt{\lambda}$ in accordance with \cite{Hatta:2007he}.

At this point, there is a valid comment to be made. Holography usually works in the $ N_c \to \infty $
limit. In order to obtain corrections with a finite number of colors one has to consider quantum corrections to the gravitational action. For instance, Wilson loops with finite $ N_c $ corrections have been proposed in \cite{Faraggi:2011bb,Buchbinder:2014nia,Faraggi:2014tna}.  

In this work we are considering a simple AdS/QCD model with  large $N_c$. But, from a phenomenological point of view, inspired in QCD results, we introduce a factor $1/N_c^2$ assuming a finite $N_c$ and rewrite (\ref{difu_1}), to obtain the scattering amplitude:
\begin{equation}\label{tb}
T(\beta, \tau) = \int_{0}^{\infty} d\beta' \;T_0 (\beta, \beta', \tau) \;|\Phi(\beta')|^2,
\end{equation}

\noindent where 
\begin{equation}\label{t0}
T_0 (\beta, \beta', \tau) \equiv \frac{1}{N_c^2} \frac{e^{\omega_0 \tau}}{2 \sqrt{\pi D \tau}} e^{- \frac{(\beta - \beta' )^2}{4 D \tau}}.
\end{equation}

Then, using the unitarity bound limit $T_0 (\beta, \beta', \tau) \sim 1 $ and neglecting the effects of pre-factor $1/ (2 \sqrt{\pi D \tau})$ in (\ref{t0}) one gets:
\begin{eqnarray}
e^{\omega_0 \tau}  e^{- \frac{(\beta - \beta' )^2}{4 D \tau}} & = & N_c^2 \nonumber \\
(\beta - \beta' )^2 & = & 4D \tau \left[ -\ln N_c^2 + \omega_0 \tau \right]  \nonumber
\end{eqnarray}

\noindent and finally to $\beta = \beta_s(\beta', \tau) $
\begin{equation}\label{bs}
\beta_s(\beta', \tau) = \beta' + \sqrt{4D \tau \left[ -\ln N_c^2 + \omega_0 \tau \right]},
\end{equation}

\noindent where one can define $\beta_{s0}(\tau) \equiv \sqrt{4D \tau \left[ -\ln N_c^2 + \omega_0 \tau \right]}$.

The above result can be understood as the elementary scattering amplitude between states located around $\beta$ and $\beta'$ in the single-pomeron-exchange approximation. In this scenario, the unitary bound determines the saturation line for a dilaton localized at $\beta'$, so one can write:
\begin{equation}
\beta_s(\beta', \tau) = \beta' + \beta_{s0}(\tau)
\end{equation}

\noindent where $\beta_{s0}(\tau)$ refers to a dilaton located at $u \sim u_0$, with $u_0 = \Lambda R^2$. Using (\ref{bs}), one can infer the critical value of the parameter $\tau$, given by $\tau_{cr} = (1/\omega_0) \ln N_c^2$.

The expression (\ref{bs}) is valid for the regime
\begin{equation}\label{cond}
\tau_{cr} < \tau < \tau_c, \;\;\;\; \textrm{with} \; \;\;\; \tau_c - \tau_{cr}\simeq 2 \ln N^2_c/ \sqrt{\lambda}
\end{equation}

Using the single pomeron approximation, one can compute Eq. (\ref{tb}) with corrections for unitarity violations, such that for $\tau> \tau_{cr} $, one can write 
\begin{equation}\label{cuv}
T(\beta, \tau) \simeq \int_{0}^{\beta - \beta_{s0}} d\beta' \frac{1}{N_c^2} e^{\omega_0 \tau} e^{- \frac{(\beta - \beta' )^2}{4 D \tau}}e^{- \Delta \beta'} + \int_{\beta - \beta_{s0}}^{\infty} d\beta' e^{- \Delta \beta'}
\end{equation}

Although Eq. (\ref{cuv}) was defined for weak amplitude regime, i.e., $T(\beta, \tau)\ll1$, one can  extrapolate its use to the vicinity of the unitarity limit $T\sim 1$ in the same sense as discussed in \cite{Hatta:2007he} . Due to this, the second integral in (\ref{cuv}) refers to the dilaton contribution for large $u$ values  and has a trivial solution. The first integral in (\ref{cuv}) contains the dilaton contribution for small values of $u$ which are weakly scattered for the ${\cal R}-$bosons. The overall amplitude is dominated  by components living the infrared cutoff, as well as in the perturbative QCD.

The integrand of the first integral becomes maximum for $\beta'  = \beta - 2 \Delta D \tau$, but for $\beta < 2 \Delta D \tau$ the integral is dominated by its lower limit $\beta' =0$, then
\begin{equation}\label{cuv_1}
T(\beta, \tau) \sim \frac{1}{N_c^2} e^{\omega_0 \tau} e^{- \frac{\beta^2}{4 D \tau}} + \frac{1}{\Delta} e^{- \Delta (\beta - \beta_{s0})}.
\end{equation}

\noindent 
For $\beta_s \simeq \beta_{s0}(\tau)$ the amplitude in (\ref{cuv_1}) becomes of ${\cal O}(1)$.

The calculations performed in (\ref{cuv_1})  were based on the condition $\beta < 2 \Delta D \tau$, which implies in $\beta_s < 2 \Delta D \tau$ or $\beta_{s0} < 2 \Delta D \tau$. Then, using (\ref{bs}), one can define

\begin{equation}\label{td}
\tau_d \equiv \frac{\ln N^2_c}{\omega_0 - D \Delta^2} = \frac{\ln N^2_c}{1 -(2/ \sqrt{\lambda})(\Delta^2+1)}.
\end{equation}

\noindent 
For values of $\tau < \tau_d$ the second term in (\ref{cuv_1}) is suppressed by the first, then

\begin{equation}\label{cuv_2}
T(\beta, \tau) \simeq \frac{1}{N_c^2} e^{\omega_0 \tau} e^{- \frac{\beta^2}{4 D \tau}} 
\end{equation}

\noindent with $\beta < 2 \Delta D \tau$. 
In this regime, the scattering is governed by dilaton component near the infrared cutoff, given by $u_0 = \Lambda R^2$.

For values of $\beta > 2 \Delta D \tau$, still keeping $\tau < \tau_d$, one gets 
\begin{equation}\label{cuv_3}
T(\beta, \tau) \simeq \frac{1}{N_c^2} e^{(\omega_0 +D\Delta^2)\tau - \Delta\beta}
\end{equation}

The amplitude in (\ref{cuv}) can be replaced by (\ref{cuv_2}) because the expression found in (\ref{cuv_1}) can not be considered for large values of $\tau$ and $\beta$. Then, to take into account the unitarity bound, one can write
\begin{equation}\label{sist}
T(\beta, \tau) \simeq  \left\{ 
\begin{array}{rrr} 
\frac{1}{N_c^2} e^{\omega_0 \tau} e^{- \frac{\beta^2}{4 D \tau}}, & \textrm{if} & \tau < \tau_s (\beta) \\
1,                                                                                                          & \textrm{if}  &\tau \gtrsim \tau_s(\beta)
\end{array}
\right.
\end{equation}

\noindent where $\tau_s(\beta)$ is the saturation line expressed as a function of $\beta$. One can also understand $\tau_s(\beta)$ as the rapidity at which the unitary corrections become important on the resolution scale fixed by $\beta$. 
From (\ref{bs}) one can check that
\begin{equation}\label{taus}
\tau_s(\beta) = \frac{\ln N^2_c}{2 \omega_0} + \sqrt{\left( \frac{ln N^2_c}{2 \omega_0} \right)^2 + \frac{\beta^2}{4 D \omega_0} }.
\end{equation}

From the set of equations (\ref{bs}), (\ref{td}), (\ref{sist}), (\ref{taus}) and the condition (\ref{cond}) one can plot a phase diagram in the kinematical plane $\tau - \beta $, with $\tau = \ln(s/Q^2)$ and $\beta = \ln (Q^2/\Lambda^2)$, for a deep inelastic scattering in the high energy limit and strong coupling when $\ln N^2_c \gg \sqrt{\lambda}$.

These results are in agreement with those found in \cite{Hatta:2007he} for the saturation line using the hardwall model.

\section{Conclusions}

The present work was based on the use of the AdS/CFT correspondence to circumvent the difficulty of studying physical phenomena outside the perturbative regime of QCD. In addition, this work can be seen and understood through two main goals. The first one is the calculation of the DIS structure functions at exponentially small $x$, shown in section III, and the second one is a study of saturation line, shown in section V. In both cases, we used the holographic approach of the softwall model in ten dimensions. A comparison between the structure functions from the soft and hardwall models was given in section IV. 

The exponentially small $x$ regime is very interesting because it implies non-local modifications of the structure functions with the inclusion of a power differential operator in the $\tilde{t}$ channel, in ten dimensions. This Regge factor is not trivial because although the four-dimensional t vanishes, the ten-dimensional $\tilde{t} $ is not zero, including derivatives in transverse directions. Although $\tilde{t}$ is small in scale strings of the order of $ (g_s N)^{- 1/2} $, its effects become large for  exponentially large $ \tilde{s} $ or equivalently to exponentially small $ x $.

The choice of the softwall model was due to the fact that this model provides linear Regge trajectories, unlike the hardwall.

Another question that can be analyzed from the DIS in the exponentially small $x$ regime, and that is the second objective of this work, concerns the study of the saturation line.

In the work \cite{Hatta:2007he} the hardwall model  was used in the exponentially small $x$ regime  within the DIS, to study the saturation line obtaining a phase diagram similar to the one expected for QCD.

The study made in this present work aimed to investigate if the introduction of a dilatonic scalar field
in the the gauge fields, which is a softwall approach requirement, produces some  modification in the phase diagram found in \cite{Hatta:2007he}. By the arguments used in Section V, the softwall gauge fields can be perfectly used for this study and we found a similar form of the structure functions $W_2(x, q^2)$. 

We have also seen in section IV  that the DIS formulations of hardwall and softwall are equivalent up to exponentially small terms thus  supporting the results found in \cite{Polchinski:2002jw} and  \cite{Hatta:2007he}. In particular, this can be seen explicitly in Eqs. (\ref{f1011}) and (\ref{diff2}) where the softwall contributions are expressed in terms of hardwall contributions plus corrections of the order of $x^{\Delta + n }$, with $\Delta, n > 0$. So, the difference between the structure functions of these models at exponentially small $x$ are negligible. Then, we conclude that the IR difference between these two models are not important for exponentially small $x$ physics. 

Finally, it would be interesting to investigate if corrections of the order of $1/N_c $ could modify these results. A possible way to include these corrections may be inspired in the works 
\cite{Faraggi:2011bb,Buchbinder:2014nia,Faraggi:2014tna} where $1/N_c$ terms were considered in the calculation of Wilson loops. 

\begin{acknowledgments}

The authors are partially supported by CAPES, CNPq and FAPERJ, Brazilian agencies.
 
\end{acknowledgments}


\begin{thebibliography}{99}


\bibitem{Manohar:1992tz} 
  A.~V.~Manohar,
  ``An Introduction to spin dependent deep inelastic scattering,''
  In *Lake Louise 1992, Symmetry and spin in the standard model* 1-46
  [hep-ph/9204208].
\bibitem{Maldacena:1997re} 
  J.~M.~Maldacena,
  ``The Large N limit of superconformal field theories and supergravity,''
  Adv.\ Theor.\ Math.\ Phys.\  {\bf 2}, 231 (1998)
  [hep-th/9711200].
\bibitem{Gubser:1998bc} 
  S.~S.~Gubser, I.~R.~Klebanov and A.~M.~Polyakov,
  ``Gauge theory correlators from noncritical string theory,''
  Phys.\ Lett.\ B {\bf 428}, 105 (1998)
  [hep-th/9802109].
\bibitem{Witten:1998qj} 
  E.~Witten,
  ``Anti-de Sitter space and holography,''
  Adv.\ Theor.\ Math.\ Phys.\  {\bf 2}, 253 (1998)
  [hep-th/9802150].
\bibitem{Witten:1998zw} 
  E.~Witten,
  ``Anti-de Sitter space, thermal phase transition, and confinement in gauge theories,''
  Adv.\ Theor.\ Math.\ Phys.\  {\bf 2}, 505 (1998)
  [hep-th/9803131].
\bibitem{Aharony:1999ti} 
  O.~Aharony, S.~S.~Gubser, J.~M.~Maldacena, H.~Ooguri and Y.~Oz,
  ``Large N field theories, string theory and gravity,''
  Phys.\ Rept.\  {\bf 323}, 183 (2000)
  [hep-th/9905111].
\bibitem{Polchinski:2001tt} 
  J.~Polchinski and M.~J.~Strassler,
  ``Hard scattering and gauge / string duality,''
  Phys.\ Rev.\ Lett.\  {\bf 88}, 031601 (2002)
  [hep-th/0109174].
\bibitem{BoschiFilho:2002vd} 
  H.~Boschi-Filho and N.~R.~F.~Braga,
  ``Gauge / string duality and scalar glueball mass ratios,''
  JHEP {\bf 0305}, 009 (2003)
  [hep-th/0212207].
\bibitem{BoschiFilho:2002ta} 
  H.~Boschi-Filho and N.~R.~F.~Braga,
  ``QCD / string holographic mapping and glueball mass spectrum,''
  Eur.\ Phys.\ J.\ C {\bf 32}, 529 (2004)
  [hep-th/0209080].
\bibitem{deTeramond:2005su} 
  G.~F.~de Teramond and S.~J.~Brodsky,
  ``Hadronic spectrum of a holographic dual of QCD,''
  Phys.\ Rev.\ Lett.\  {\bf 94}, 201601 (2005)
  [hep-th/0501022].
\bibitem{Erlich:2005qh} 
  J.~Erlich, E.~Katz, D.~T.~Son and M.~A.~Stephanov,
  ``QCD and a holographic model of hadrons,''
  Phys.\ Rev.\ Lett.\  {\bf 95}, 261602 (2005)
  [hep-ph/0501128].
\bibitem{BoschiFilho:2005yh} 
  H.~Boschi-Filho, N.~R.~F.~Braga and H.~L.~Carrion,
  ``Glueball Regge trajectories from gauge/string duality and the Pomeron,''
  Phys.\ Rev.\ D {\bf 73}, 047901 (2006)
  [hep-th/0507063].
\bibitem{Capossoli:2013kb} 
  E.~F.~Capossoli and H.~Boschi-Filho,
  ``Odd spin glueball masses and the Odderon Regge trajectories from the holographic hardwall model,''
  Phys.\ Rev.\ D {\bf 88}, no. 2, 026010 (2013)
  [arXiv:1301.4457 [hep-th]].
\bibitem{Polchinski:2002jw} 
  J.~Polchinski and M.~J.~Strassler,
  ``Deep inelastic scattering and gauge / string duality,''
  JHEP {\bf 0305}, 012 (2003)
  [hep-th/0209211].
\bibitem{Brower:2006ea} 
  R.~C.~Brower, J.~Polchinski, M.~J.~Strassler and C.~-ITan,
  ``The Pomeron and gauge/string duality,''
  JHEP {\bf 0712}, 005 (2007)
  [hep-th/0603115].
\bibitem{Hatta:2007he} 
  Y.~Hatta, E.~Iancu and A.~H.~Mueller,
  ``Deep inelastic scattering at strong coupling from gauge/string duality: The Saturation line,''
  JHEP {\bf 0801}, 026 (2008)
  [arXiv:0710.2148 [hep-th]].
\bibitem{BallonBayona:2007rs} 
  C.~A.~Ballon Bayona, H.~Boschi-Filho and N.~R.~F.~Braga,
  ``Deep inelastic structure functions from supergravity at small x,''
  JHEP {\bf 0810}, 088 (2008)
  [arXiv:0712.3530 [hep-th]].
\bibitem{Gao:2009ze} 
  J.~H.~Gao and B.~W.~Xiao,
  Phys.\ Rev.\ D {\bf 80}, 015025 (2009)
  [arXiv:0904.2870 [hep-ph]].
\bibitem{Brower:2010wf} 
  R.~C.~Brower, M.~Djuric, I.~Sarcevic and C.~I.~Tan,
  ``String-Gauge Dual Description of Deep Inelastic Scattering at Small-$x$,''
  JHEP {\bf 1011}, 051 (2010)
  [arXiv:1007.2259 [hep-ph]].
\bibitem{Karch:2006pv} 
  A.~Karch, E.~Katz, D.~T.~Son and M.~A.~Stephanov,
  ``Linear confinement and AdS/QCD,''
  Phys.\ Rev.\ D {\bf 74}, 015005 (2006)
  [hep-ph/0602229].
\bibitem{Colangelo:2007pt}
  P.~Colangelo, F.~De Fazio, F.~Jugeau and S.~Nicotri,
  ``On the light glueball spectrum in a holographic description of QCD,''
  Phys.\ Lett.\ B {\bf 652} (2007) 73
  [hep-ph/0703316].
\bibitem{BallonBayona:2007qr} 
  C.~A.~Ballon Bayona, H.~Boschi-Filho and N.~R.~F.~Braga,
  ``Deep inelastic scattering from gauge string duality in the soft wall model,''
  JHEP {\bf 0803}, 064 (2008)
  [arXiv:0711.0221 [hep-th]].
\bibitem{Braga:2011wa} 
  N.~R.~F.~Braga and A.~Vega,
  ``Deep inelastic scattering of baryons in a modified soft wall model,''
  Eur.\ Phys.\ J.\ C {\bf 72}, 2236 (2012)
  [arXiv:1110.2548 [hep-ph]].
\bibitem{BallonBayona:2010ae} 
  C.~A.~Ballon Bayona, H.~Boschi-Filho, N.~R.~F.~Braga and M.~A.~C.~Torres,
  ``Deep inelastic scattering for vector mesons in holographic D4-D8 model,''
  JHEP {\bf 1010}, 055 (2010)
  [arXiv:1007.2448 [hep-th]].
\bibitem{Cornalba:2008sp} 
  L.~Cornalba and M.~S.~Costa,
  ``Saturation in Deep Inelastic Scattering from AdS/CFT,''
  Phys.\ Rev.\ D {\bf 78}, 096010 (2008)
  [arXiv:0804.1562 [hep-ph]].
\bibitem{Pire:2008zf} 
  B.~Pire, C.~Roiesnel, L.~Szymanowski and S.~Wallon,
  ``On AdS/QCD correspondence and the partonic picture of deep inelastic scattering,''
  Phys.\ Lett.\ B {\bf 670}, 84 (2008)
  [arXiv:0805.4346 [hep-ph]].
\bibitem{Albacete:2008ze} 
  J.~L.~Albacete, Y.~V.~Kovchegov and A.~Taliotis,
  ``DIS on a Large Nucleus in AdS/CFT,''
  JHEP {\bf 0807}, 074 (2008)
  [arXiv:0806.1484 [hep-th]].
\bibitem{BallonBayona:2008zi} 
  C.~A.~Ballon Bayona, H.~Boschi-Filho and N.~R.~F.~Braga,
  ``Deep inelastic scattering from gauge string duality in D3-D7 brane model,''
  JHEP {\bf 0809}, 114 (2008)
  [arXiv:0807.1917 [hep-th]].
\bibitem{Yoshida:2009dw} 
  Y.~Yoshida,
  ``The Virtual Photon Structure Functions and AdS/QCD Correspondence,''
  Prog.\ Theor.\ Phys.\  {\bf 123}, 79 (2010)
  [arXiv:0902.1015 [hep-th]].
\bibitem{Hatta:2009ra} 
  Y.~Hatta, T.~Ueda and B.~W.~Xiao,
  ``Polarized DIS in N=4 SYM: Where is spin at strong coupling?,''
  JHEP {\bf 0908}, 007 (2009)
  [arXiv:0905.2493 [hep-ph]].
\bibitem{Avsar:2009xf} 
  E.~Avsar, E.~Iancu, L.~McLerran and D.~N.~Triantafyllopoulos,
  ``Shockwaves and deep inelastic scattering within the gauge/gravity duality,''
  JHEP {\bf 0911}, 105 (2009)
  [arXiv:0907.4604 [hep-th]].
\bibitem{Cornalba:2009ax} 
  L.~Cornalba, M.~S.~Costa and J.~Penedones,
  ``Deep Inelastic Scattering in Conformal QCD,''
  JHEP {\bf 1003}, 133 (2010)
  [arXiv:0911.0043 [hep-th]].
\bibitem{Bayona:2009qe} 
  C.~A.~B.~Bayona, H.~Boschi-Filho and N.~R.~F.~Braga,
  ``Deep inelastic scattering off a plasma with flavour from D3-D7 brane model,''
  Phys.\ Rev.\ D {\bf 81}, 086003 (2010)
  [arXiv:0912.0231 [hep-th]].
\bibitem{Cornalba:2010vk} 
  L.~Cornalba, M.~S.~Costa and J.~Penedones,
  ``AdS black disk model for small-x DIS,''
  Phys.\ Rev.\ Lett.\  {\bf 105}, 072003 (2010)
  [arXiv:1001.1157 [hep-ph]].
  
 
\bibitem{Koile:2011aa}
  E.~Koile, S.~Macaluso and M.~Schvellinger,
  ``Deep Inelastic Scattering from Holographic Spin-One Hadrons,''
  JHEP {\bf 1202} (2012) 103
  [arXiv:1112.1459 [hep-th]].
  

\bibitem{Koile:2013hba}
  E.~Koile, S.~Macaluso and M.~Schvellinger,
  ``Deep inelastic scattering structure functions of holographic spin-1
hadrons with $N_f \geq 1$,''
  JHEP {\bf 1401} (2014) 166
  [arXiv:1311.2601 [hep-th]].
  

\bibitem{Koile:2014vca}
  E.~Koile, N.~Kovensky and M.~Schvellinger,
  ``Hadron structure functions at small $x$ from string theory,''
  JHEP {\bf 1505} (2015) 001
  [arXiv:1412.6509 [hep-th]].
  
  
\bibitem{Koile:2015qsa}
  E.~Koile, N.~Kovensky and M.~Schvellinger,
  ``Deep inelastic scattering cross sections from the gauge/string
duality,''
  arXiv:1507.07942 [hep-th].
  
\bibitem{Csaki:1998qr} 
  C.~Csaki, H.~Ooguri, Y.~Oz and J.~Terning,
  ``Glueball mass spectrum from supergravity,''
  JHEP {\bf 9901}, 017 (1999)
  [hep-th/9806021].
\bibitem{Gribov:1972ri} 
  V.~N.~Gribov and L.~N.~Lipatov,
  ``Deep inelastic e p scattering in perturbation theory,''
  Sov.\ J.\ Nucl.\ Phys.\  {\bf 15}, 438 (1972)
  [Yad.\ Fiz.\  {\bf 15}, 781 (1972)],   
  
  \bibitem{DGLAP}
  G.~Altarelli and G.~Parisi,
  ``Asymptotic Freedom in Parton Language,''  Nucl.\ Phys.\ B {\bf 126}, 298 (1977) 
  
  \bibitem{DGLAP2}
  Y.~L.~Dokshitzer, ``Calculation of the Structure Functions for Deep Inelastic Scattering and e+ e- Annihilation by Perturbation Theory in Quantum Chromodynamics.,''  Sov.\ Phys.\ JETP {\bf 46}, 641 (1977)   [Zh.\ Eksp.\ Teor.\ Fiz.\  {\bf 73}, 1216 (1977)].
\bibitem{Lipatov:1976zz} 
  L.~N.~Lipatov, ``Reggeization of the Vector Meson and the Vacuum Singularity in Nonabelian Gauge Theories,''  Sov.\ J.\ Nucl.\ Phys.\  {\bf 23}, 338 (1976)  [Yad.\ Fiz.\  {\bf 23}, 642 (1976)], 
  
  \bibitem{BFKL}
  E.~A.~Kuraev, L.~N.~Lipatov and V.~S.~Fadin, ``The Pomeranchuk Singularity in Nonabelian Gauge Theories,''  Sov.\ Phys.\ JETP {\bf 45}, 199 (1977)   [Zh.\ Eksp.\ Teor.\ Fiz.\  {\bf 72}, 377 (1977)] 
  
  \bibitem{BFKL2}
    I.~I.~Balitsky and L.~N.~Lipatov,  ``The Pomeranchuk Singularity in Quantum Chromodynamics,''   Sov.\ J.\ Nucl.\ Phys.\  {\bf 28}, 822 (1978)   [Yad.\ Fiz.\  {\bf 28}, 1597 (1978)].
\bibitem{Kovchegov:1999yj} 
  Y.~V.~Kovchegov, ``Small x F(2) structure function of a nucleus including multiple pomeron exchanges,''  Phys.\ Rev.\ D {\bf 60}, 034008 (1999)   [hep-ph/9901281] 
  
  \bibitem{BK}
  Y.~V.~Kovchegov, ``Unitarization of the BFKL pomeron on a nucleus,'' Phys.\ Rev.\ D {\bf 61}, 074018 (2000)   [hep-ph/9905214].

\bibitem{JalilianMarian:1997jx} 
J.~Jalilian-Marian, A.~Kovner, A.~Leonidov and H.~Weigert, ``The BFKL equation from the Wilson renormalization group,'' Nucl.\ Phys.\ B {\bf 504}, 415 (1997) [hep-ph/9701284]

\bibitem{JIKLMW2}
J.~Jalilian-Marian, A.~Kovner, A.~Leonidov and H.~Weigert, ``The Wilson renormalization group for low x physics: Towards the high density regime,''  Phys.\ Rev.\ D {\bf 59}, 014014 (1998) [hep-ph/9706377]

\bibitem{JIKLMW3}
  J.~Jalilian-Marian, A.~Kovner, A.~Leonidov and H.~Weigert, ``Unitarization of gluon distribution in the doubly logarithmic regime at high density,''  Phys.\ Rev.\ D {\bf 59}, 034007 (1999) [Phys.\ Rev.\ D {\bf 59}, 099903 (1999)]  [hep-ph/9807462], 
  
  \bibitem{JIKLMW4}
  A.~Kovner, J.~G.~Milhano and H.~Weigert, ``Relating different approaches to nonlinear QCD evolution at finite gluon density,''  Phys.\ Rev.\ D {\bf 62}, 114005 (2000)  [hep-ph/0004014], 
  
  \bibitem{JIKLMW5}
  E.~Iancu, A.~Leonidov and L.~D.~McLerran, ``Nonlinear gluon evolution in the color glass condensate. 1.,'' Nucl.\ Phys.\ A {\bf 692}, 583 (2001) [hep-ph/0011241], 
  
  \bibitem{JIKLMW6}
  E.~Iancu, A.~Leonidov and L.~D.~McLerran, ``The Renormalization group equation for the color glass condensate,'' Phys.\ Lett.\ B {\bf 510}, 133 (2001) [hep-ph/0102009], 
  
  \bibitem{JIKLMW7}
  E.~Ferreiro, E.~Iancu, A.~Leonidov and L.~McLerran, ``Nonlinear gluon evolution in the color glass condensate. 2.,'' Nucl.\ Phys.\ A {\bf 703}, 489 (2002)  [hep-ph/0109115] 
  
  \bibitem{JIKLMW8}
  E.~Iancu and L.~D.~McLerran, ``Saturation and universality in QCD at small x,'' Phys.\ Lett.\ B {\bf 510}, 145 (2001) [hep-ph/0103032].

\bibitem{Faraggi:2011bb} 
  A.~Faraggi and L.~A.~Pando Zayas,
  ``The Spectrum of Excitations of Holographic Wilson Loops,''
  JHEP {\bf 1105}, 018 (2011)
  [arXiv:1101.5145 [hep-th]].
\bibitem{Buchbinder:2014nia} 
  E.~I.~Buchbinder and A.~A.~Tseytlin,
  ``1/N correction in the D3-brane description of a circular Wilson loop at strong coupling,''
  Phys.\ Rev.\ D {\bf 89}, no. 12, 126008 (2014)
  [arXiv:1404.4952 [hep-th]].
\bibitem{Faraggi:2014tna} 
  A.~Faraggi, J.~T.~Liu, L.~A.~Pando Zayas and G.~Zhang,
  ``One-loop structure of higher rank Wilson loops in AdS/CFT,''
  Phys.\ Lett.\ B {\bf 740}, 218 (2015)
  [arXiv:1409.3187 [hep-th]].





\end{thebibliography}
\end{document}